\documentclass[12pt]{article} 

\usepackage{amsfonts}
 \usepackage{amssymb}
 \usepackage{amsmath}

\def\pmx{\begin{pmatrix}}
\def\emx{\end{pmatrix}}
\def\bsq{\begin{subequations}}
\def\esq{\end{subequations}}

\newtheorem{thm}{Theorem}

\def\be{\begin{eqnarray}}
\def\ee{\end{eqnarray}}
\def\bee{\begin{eqnarray*}}
\def\eee{\end{eqnarray*}}
\def\ds{\displaystyle}
\def\bra{\langle}
\def\ket{\rangle}
\def\dg{\dagger}
\def\kb{ \ket \bra }

\def\raw{\rightarrow}

\def\half{{\textstyle \frac{1}{2}}}

\def\thrd{{\textstyle \frac{1}{3}}}
 \def\tr{\hbox{Tr} \,}
 \def\trp{\hbox{Tr}}
\def\mm{ \! - \! }

\def\nn{\nonumber}
\def\ot{\otimes}
\def\op{\oplus}

\def\wh{\widehat}
\def\dep{{\rm dep}}
\def\hv{{\rm Holv}}
\def\av{{\rm av}}
\def\bw{{ \bf w}}
\def\mm{ \! - \! }
\def\pp{ \! + \! }
\def\dtsig{\cdot{\mathbf \sigma}}
\def\cd{{\mathcal D}}
\def\qed{\qquad{\bf QED}}
\def\pf{ \noindent{\bf Proof:} }

\allowdisplaybreaks[1]

\newcommand{\proj}[1]{ | #1 \kb  #1|}
\newcommand{\norm}[1]{ \| #1  \|}
\newcommand{\bgnorm}[1]{ \big\| #1  \big\|}

\title{Maximal output purity and capacity for   \\  asymmetric unital qudit channels}

    \author{Nilanjana Datta \\ Statistical Laboratory \\
    Centre for Mathematical Sciences \\
University of Cambridge \\
Wilberforce Road,
Cambridge,
CB3 0WB UK \\
    {\small N.Datta@statslab.cam.ac.uk}
\and Mary Beth Ruskai \thanks{Partially supported  by
 the National Security Agency (NSA) and
 Advanced Research and Development Activity (ARDA) under
Army Research Office (ARO) contract number 
     DAAD19-02-1-0065, and by the National Science
        Foundation under Grant  DMS-0314228.}
      \\ Department of Mathematics \\
Tufts University \\
     Medford, MA 02155 USA\\ 
    {\small  Marybeth.Ruskai@tufts.edu}}

\begin{document}

\maketitle

  \begin{abstract}
  We consider generalizations of depolarizing channels to maps
  of the form  $   \Phi(\rho) = \sum_k a_k V_k \rho V_k^{\dg} + (1-a) (\tr \rho) \, \tfrac{1}{d} I $
  with $V_k$ unitary and $ \sum_k a_k = a < 1$.
We show that one can construct unital channels of this type for  which
the input which achieves maximal output purity is unique.   We give
conditions on $V_k$ under which multiplicativity of the maximal $p$-norm
and additivity of the minimal output entropy can be proved 
for $\Phi \ot \Omega$
with $\Omega$ arbitrary.    We also show that the Holevo capacity need not
equal $\log d - S_{\min}(\Phi)$ as one might expect for a convex
combination of unitary conjugations.    
  \end{abstract}

  \section{Introduction}
  
 The depolarizing channel   $\Gamma_a^\dep$ has the form
 \be  \label{dep}
     \Gamma_a^\dep(\rho) = a \rho + (1-a) (\tr \rho) \, \tfrac{1}{d} I .
  \ee
with $- \frac{1}{d^2 -1} \leq a \leq 1$.
In this paper, we consider channels of the more general form
  \be  \label{chandef}
     \Phi(\rho) = \sum_k a_k V_k \rho V_k^{\dg} + (1-a) (\tr \rho) \, \tfrac{1}{d} I
  \ee
 with $0 < a_k$,  $0 < a = \sum_k a_k  < 1$ and $V_k$ unitary.  
 
  We describe and study several subclasses of these  channels \eqref{chandef}, 
  showing that they can exhibit  different types of behavior.    
 Those with simultaneously diagonal $V_k$ have a high level of symmetry
  and much in common with depolarizing channels.    However, we also
  construct asymmetric channels with a unique state of minimal output entropy
  and other behavior more typical of non-unital channels; although additivity
  can be proved for the minimal output entropy, this does not imply additivity
  of the capacity because the optimal average output is not $\tfrac{1}{d} I$.
  
  This paper is organized as follows. Section~\ref{sect:back} contains
 some terminology and notation as well as considerable background 
 material on various types of channels and their behavior. 
  In Section~\ref{sect:results}, we state and prove some theorems about
 minimal output purity for the channels we consider. In 
  Section~\ref{sect:diag} we consider a special subclass of channels 
  which satisfy \eqref{chandef} and exhibit behavior similar to unital
  qubit channels.  In
 Section~\ref{sect:exam}, which is the heart of the paper, we describe 
several  types of asymmetric channels to which our results 
 can be applied.     In Section~\ref{sect:num} we report the results of
 numerical tests on channel capacity.    
    
 \section{Background}  \label{sect:back}
 
 \subsection{General notation and terminology}
 
 We restrict attention to finite dimensional spaces ${\bf C}^d$ 
 and denote the space of $d \times d $ complex 
matrices as $M_d= {\cal{B}}({\bf{C}}^d)$.
 By a channel $\Phi$ we mean a  completely positive, trace preserving
 (CPT) map $\Phi: M_d \mapsto M_d$.   Let $\cd = \{ \rho : \rho \geq 0, \tr \rho = 1\}$ 
 denote the set of density matrices in $M_d$.   Let $S(\gamma) = - \tr \gamma \log \gamma$
 denote the quantum entropy of a state $\gamma \in \cd$.
 For a CPT map $\Phi$, one can define  the maximal output
 $p$-norm
 \be  \label{nup}
    \nu_p(\Phi) = \sup_{\gamma \in \cd} \norm{ \Phi(\gamma) }_p,
 \ee
the minimal output entropy
  \be  \label{Smin}
   S_{\min}(\Phi) = \inf_{\rho \in \cd} S[\Phi(\rho)],
 \ee
and the  Holevo capacity
 \be  \label{hvcap}
    C_\hv(\Phi) = \sup_{\{ \pi_j, \rho_j \} }  \Big( S[\Phi(\rho_\av)] - \sum_j \pi_j S[\Phi(\rho_j)] \Big),
 \ee
 where $\rho_\av = \sum_j \pi_j \rho_j$, and the supremum is taken over 
 all ensembles $\{ \pi_j, \rho_j \}$ with $\rho_j \in \cd$, $\pi_j > 0$ and $\sum_j \pi_j = 1$.
 Both $ S_{\min}(\Phi) $ and $ C_\hv(\Phi) $ are conjectured to be additive
 over tensor products, i.e., to satisfy
 \be  \label{Sadd}
   S_{\min}(\Phi \ot \Omega) & = &    S_{\min}(\Phi)  + S_{\min}(\Omega), \qquad {\rm and} \\
    C_\hv(\Phi \ot \Omega) & = &     C_\hv(\Phi)  + C_\hv(\Omega)   \label{Cadd}
 \ee
 Shor  showed  \cite{ShorEQ} that these conjectures (and several related ones) are 
 equivalent in the global sense that both are either true for all general channels 
 $\Phi : M_d \mapsto M_n$ or both are false.   However, they are not necessarily
 equivalent for  individual channels, and we will study them separately for the
  examples in this paper.
 
  Shor also proved \cite{ShorEBT}  that both \eqref{Sadd} and  \eqref{Cadd}
hold   for entanglement breaking    (EB) channels.    King \cite{King4} gave
an alternative proof based on multiplicativity of $ \nu_p(\Phi)$.
 A CP map $\Phi$ is    EB if $(I \ot \Phi)(\rho)$ is separable
for all input states $\rho$.    A CPT map which is also EB is denoted as EBT.     It
was shown in \cite{HSR} that a CP map is EB if all its Kraus operators can
be chosen to have rank one, or if  $(I \ot \Phi)(\proj{\Psi})$ is separable for some
maximally entangled $|\Psi\rangle$.  Any EBT channel be written as 
\be
   \Phi(\rho) = \sum_k  \gamma_k \tr \rho E_k,
\ee
with $\{ E_k \}$ a POVM, and each $\gamma_k \in {\cal D}$.  When 
$\{ |e_k \ket \}$ is an orthonormal basis for ${\bf C}^d$ and $E_k = \proj{e_k}$
the channel is called CQ  (classical-quantum); and when each  
$\gamma_k = \proj{e_k}$ it is called QC (quantum-classical).

 The following max-min characterizations of $C_\hv(\Phi)$ in terms
 of the relative entropy $H(\rho,\gamma) = \tr \rho (\log \rho - \log \gamma)$
  are extremely useful.    They were obtained  
   independently in \cite{OPW} and \cite{SW3}.
\bsq  \label{maxmin} \be 
     C_\hv(\Phi) & = & \inf_{\gamma \in \cd} \sup_{\omega \in \cd} 
           H\big[ \Phi(\omega), \Phi(\gamma) \big] \\   \label{maxmin2}
         & = &   \sup_{\omega \in \cd} H\big[ \Phi(\omega), \Phi(\rho_\av) \big]  \\
        & = &    H\big[ \Phi(\rho_j), \Phi(\rho_\av) \big],   \label{maxmin3}
 \ee \esq
 where $\rho_\av$ is the optimal
 average input and $\rho_j$ is any input in the optimal signal ensemble.  It can be
 shown \cite{ERATO} that \eqref{maxmin2} and \eqref{maxmin3} are equivalent to the
 statement that the points $\big(\rho_i, S(\rho_i) \big)$ define a supporting
 hyperplane for the convex optimization problem \eqref{hvcap}.

 \subsection{Depolarizing channels}
  
 The properties of the depolarizing channel are well-known  and can be
 summarized as follows.
 \begin{thm}   \label{thm:dep}
 The depolarizing channel \eqref{dep} satisifies
 \begin{enumerate}
 \renewcommand{\labelenumi}{\theenumi}
    \renewcommand{\theenumi}{\alph{enumi})}
 
 \item $\Gamma_a^\dep(I)$ is unital, i.e., $\Gamma_a^\dep(I) = I$.
 
 \item The output $\Gamma_a^\dep \big( \proj{\psi} \big)$ for any pure state
$\proj{\psi}$ has eigenvalues $[ a + \tfrac{1-a}{d}, \, \tfrac{1-a}{d}, \ldots \tfrac{1-a}{d}]$.
 
 \item  For any CPT map $\Omega$,
 $  \nu_p(\Gamma_a^\dep \ot \Omega)  =  \nu_p(\Gamma_a^\dep)  \nu_p(\Omega)   \quad \forall ~~ p  \geq 1  $.  

 \item  For any CPT map $\Omega$,      
     $S_{\min}(\Gamma_a^\dep  \ot \Omega) =   S_{\min}(\Gamma_a^\dep)   + S_{\min}(\Omega) $.
     
    \item $C_\hv(\Gamma_a^\dep) = \log d - S_{\min}(\Gamma_a^\dep) $.
        
    \item The capacity $C_\hv(\Gamma_a^\dep) $ can be achieved using $d$
    orthogonal input states.
    
    \item The optimal average input is $\tfrac{1}{d} I$.
    
    \item  For any CPT map $\Omega$,      
     $C_\hv(\Gamma_a^\dep  \ot \Omega) =   C_\hv(\Gamma_a^\dep)   + C_\hv(\Omega) $
     
     \item When $a \leq \frac{1}{d+1}$,  the channel $\Gamma_a^\dep $ is EBT.

 \end{enumerate}
 \end{thm}
 The mutiplicativity (c) was proved by King \cite{King3} for any depolarizing map, 
 including those with negative $a$; he also showed that properties (d) and (h) follow.
 Properties (d) and (h) were proved independently by Fujiwara and  Hashizum\'{e} \cite{FH}
 for maps with $a > 0$ and $\Omega =  \Gamma_a^\dep$; they 
 used a majorization argument which also implies (c).   Properties (a), (b) and (e) are
 well-known and easily verified.  Property (j)
 can be verified by computing the Choi matrix $(I \ot \Gamma_a^\dep)(\proj{\beta})$
 for a maximally entangled state $|\beta\ket$ and using Theorem~4 of \cite{HSR}.
    
It is useful to introduce the generalized Pauli operators $X_d$ and $Z_d$
defined on the standard basis so that   
$X_d |e_\ell \ket =  |e_{\ell +1} \ket$ with the addition in the subscript 
taken $\mod d$ and
$Z_d |e_\ell \ket = e^{2 \pi i \ell /d} $.   Then for any $d \times d$ 
matrix $A$,
\be   \label{noisI}
\tfrac{1}{d^2} \sum_{m = 0}^{d-1} \sum_{n = 0}^{d-1} 
     X_d^m Z_d^n A  (Z_d^{\dg})^n (X_d^{\dg})^m
       =  (\tr A) \tfrac{1}{d} I,
  \ee
and
\be  \label{chanCort}
    \Gamma_a^\dep(\rho) = \Big[a + \tfrac{1-a}{d^2} \Big] I \rho I + (1-a)
       \tfrac{1}{d^2} \underset{m, n \neq 0,0}{ \sum^{d-1}_{m=0} \sum^{d-1}_{n=0}}
     X_d^m Z_d^n \, \rho \,  (Z_d^{\dg})^n (X_d^{\dg})^m.
\ee 
Cortese \cite{C} considered channels of the form
\be   \label{Cort}
   \Phi(\rho) =  \sum_{m = 0}^{d-1} \sum_{n = 0}^{d-1}  c_{mn}
     X_d^m Z_d^n \, \rho \,   (Z_d^{\dg})^n (X_d^{\dg})^m
\ee  
 with $c_{mn} \geq  0$ and $\sum_{mn} c_{mn} = 1$, and
 showed that
 \be \label{capd}
   C_\hv(\Phi) = \log d - S_{\min}(\Phi).
 \ee
 A simplified proof of this result was given by Holevo \cite{Hv}, who  
 showed that \eqref{capd} holds for channels satisfying the covariance
  condition 
   \be  \label{Hvcov}
    \Phi(U_g \rho U_g^\dag) = 
           U_g^{\prime} \Phi(\rho) [U_g^{\prime}]^\dag \qquad \forall ~g \in {\cal G}
  \ee
when $\{U_g\}$ and $\{U_g^{\prime} \}$ are irreducible representations
of a group ${\cal G}$.   The case \eqref{Cort}  is called ``Weyl covariance''.

By using   \eqref{noisI} to rewrite the second term in (2) and the fact
that $\sum_k a_k = a$, one sees that such channels
  can be expressed as a convex combination of unitary conjugations.
We   write them in the form (2) because we exploit  their relationship
to the depolarizing channel.
However,  \eqref{capd}  need not hold for all channels of
the form \eqref{chandef}; in Sections~\ref{sect:exam} we give examples
which show that they can exhibit very different behavior.  

\subsection{Qubit channels}

As discussed in Appendix~\ref{app:qubit}, a unital qubit channel
can be written (after rotation of bases) \cite{KR1} as
 \be   \label{qubit1}
\Phi(\rho)  = \sum_{k = 0}^3 \alpha_k  \, \sigma_k \, \rho \, \sigma_k .
\ee 
It is also useful to recall that any qubit density matrix can be written as
$\rho = \half\big[ I + \bw \cdot{{\mathbf{\sigma}}}]$, where 
${\mathbf{\sigma}}$ denotes the vector of Pauli matrices and 
$\bw \in {\mathbf{C}}^3$; then the channel \eqref{qubit1} can be 
written as 
\be  \label{qubout}
\Phi(\rho) =  \half\big[ I +  \sum_{j=1}^3 \lambda_j w_j \sigma_j\big].
\ee
The relations between the parameters $\{\alpha_k\}$ and $\{\lambda_j\}$ 
are discussed
 in Appendix~\ref{app:qubit}. 

The following theorem was proved by King in \cite{King2}.

\begin{thm}  \label{thm:unit}
Let $\Phi$ be a unital qubit channel and 
$\ds{a = \max_{k=1,2,3}  |\lambda_k| = \max_{i \neq j \in 0,1,2,3} \alpha_i + \alpha_j}$.   Then
parts (c) to (h) of Theorem~\ref{thm:dep} hold, with $\Gamma_a^{\dep} $ replaced by
$\Phi$.   In addition, for those
$k$ with $|\lambda_k| = a$, the inputs $\half[I \pm \sigma_k]$ yield outputs
with eigenvalues $\half (1 \pm a)$ and, hence, have the same entropy 
as the corresponding qubit depolarizing channel. 
\end{thm}
This implies that all unital qubit channels  for which the image ellipsoid of the Bloch
sphere touches, but lies within, the sphere of
radius $a$ (which is the image of a depolarizing channel)  have the same capacity
and minimal output entropy behavior.
A unital qubit channel is EBT  \cite{EBT2} if and only if $ \sum_k |\lambda_k| \leq 1$ 
or, equivalently, if $\alpha_k \leq \half$ for all $k$.

A non-unital qubit channel can be written  (after rotation of bases) \cite{KR1}
in the form
\be  \label{qnon}
  \Phi : \half\big[ I + \bw \dtsig] \mapsto   
        \half\big[ I +  \sum_{k=1}^3 (t_k + \lambda_k w_k) \sigma_k\big].
\ee
The conditions imposed on $t_k$ and $\lambda_k$ by the CPT requirement
are given in \cite{RSW} and summarized in \cite{EBT2}.   (The special case
$t_1 = t_2 = 0$ was considered earlier in \cite{FA}.)
One expects  the generic behavior of   non-unital qubit channels
to be quite different from that of unital ones.    
 \begin{enumerate}
 \renewcommand{\labelenumi}{\theenumi}
    \renewcommand{\theenumi}{\Alph{enumi})}
 
 \item  Non-unital qubit channels typically have a unique state of optimal
 output purity.   This always holds when $t_k \neq 0$ in the direction for which 
  the ellipsoid axis   $|\lambda_k| $ is longest.

If $t_k \neq 0$ only in direction(s) orthogonal to the longest axis,
then one typically has  two
non-orthogonal states of optimal output purity (although these can
coalesce into one, as for extreme amplitude damping channels,
and can come from orthogonal inputs for a CQ channel) \cite{Fuchs,KNR1}.

 \item   $  C_\hv(\Phi) <  \log d - S_{\min}(\Phi) $
   for all non-unital qubit maps.

\item In general, the capacity $C_\hv(\Phi) $ can not be achieved using $d$
    orthogonal input states \cite{Fuchs,ERATO,KNR1,SW3}.    
    
    There are, however, a number of exceptions.  Two of these are  CQ maps which take
      $ \half\big[ I + \bw \dtsig] \mapsto   
        \half\big[ I +  t_1 \sigma_1 + \lambda_3 w_3 \sigma_3 \big]$ and QC maps 
   which take      $ \half\big[ I + \bw \dtsig] \mapsto   
        \half\big[ I +  (t_3 + \lambda_3 w_3) \sigma_3 \big]$.   The QC channels are 
        included in the larger class of channels for which $t_k \neq 0$ only 
        for the largest $|\lambda_k| $; then  $C_\hv(\Phi) $ can be achieved  
        with a pair of orthogonal inputs \cite{FN,KNR1}.
               
        \item  Properties    (c), (d), and (h) of Theorem 1 are conjectured to hold
        for non-unital qubit maps; however, a proof is known only for (c) in the case $p= 2$.
     
     \end{enumerate}
     
     \subsection{Some channels for $d > 2$}  \label{sect:othex}
     
     When $\Phi$ maps a larger space into qubit density matrices, it is possible to have 
  $  C_\hv(\Phi) =  \log d - S_{\min}(\Phi) $, even when the optimal input
  $\rho_\av \neq \frac{1}{d}I$.
 This is the case for Shor's  extended channel in  Section 9 of \cite{ShorEQ}.   
 In that case, the original map  $\Phi$ is extended to $\Phi_{\rm ext}$ for which
the optimal average input is $R_\av = \rho_{\min} \ot  \frac{1}{d^2}I $, with 
$ \rho_{\min} $ achieving  $S_{\min}(\Phi)$ for the original channel.
 Then     $\Phi_{\rm ext}(R_\av ) = \frac{1}{d} I$.   Note that one also has
 $\Phi_{\rm ext}(I_d \ot I_{d^2}  ) = I_d$ so that  $ \Phi_{\rm ext}$ is unital.
 Moreover, if $S_{\min}(\Phi)$  is achieved for more than one state,
 then the optimal average input is not unique, although the optimal average
 output is unique.

For qubits, a channel is unital if and only if it can be written as a convex
combination of unitary conjugations \cite{KR1}. It is well-known that this
result does not extend to $d > 2$.      
One well-known
example is the Werner-Holevo channel \cite{WH} for which the Kraus
operators can be written as partial isometries. This example does
satisfy \eqref{capd} as well as \eqref{Sadd} and \eqref{Cadd},
although it has only been shown to satisfy \eqref{pmult}  when
$1 \leq p \leq 2$  \cite{AF} and is known violate  \eqref{pmult}  for large $p$.

For $d = 3$, Fuchs, et al \cite{FSST} found a unital channel  which
satisfies \eqref{capd} but for which the optimal inputs are not orthogonal.
This channel is given by Eq. (19) of \cite{HSR}.

The asymmetric examples in Section~\ref{sect:exam} appear to be
the first for which a unital channel does not satisfy \eqref{capd}.

It is natural to look for classifications
of unital channels which include a type whose behavior is similar to
that of unital qubit channels. The results presented here show that
there are channels which can be written as convex
combination of unitary conjugations  which do not exhibit this behavior.
Thus we are left with the conjecture that channels of the form
\eqref{Cort} behave like unital qubit channels and, hence,
satisfy (c) to (h) of Theorem 1 with  $\Gamma$ replaced by $\Phi$, 
as in Theorem 2.

\subsection{Majorization}

We will use the notation $[x_1, x_2, \ldots x_n]  \succ  [y_1, y_2, \ldots y_n]  $
to indicate that both sets are non-negative and arranged in non-increasing order 
$x_1 \geq x_2 \geq x_3 \ldots \geq 0$ and satisfy the majorization  
condition
$\ds{\sum_{i=1}^k x_i \geq \sum_{i=1}^k y_i}$ for
$\ds{k = 1 \ldots n \! - \! 1}$ and $\ds{\sum_{i=1}^n x_i = \sum_{i=1}^n y_i}$.
   It is well-known \cite{HJ1,MO} that this implies
\be
     \sum_{j=1}^n  x_j^p \geq  \sum_{j=1}^n  y_j^p  
\ee
for all $p \geq 1$.   Therefore, whenever $\rho$ and $\gamma $ are density
matrices for which the eigenvalues of $\rho$ majorize those of $\gamma $,
$\norm{\rho}_p > \norm{\gamma}_p$ and $S(\rho) <  S(\gamma)$.

When only an inequality holds for $k = n$, we use the term
submajorize, and observe that the same conclusions follow by extending
both sets with $x_{n+1} = 0$ and $y_{n+1}$ chosen to give equality.


\section{Results on minimal output purity}
\label{sect:results}

In this section we state and prove some theorems on the minimal output purity
of certain subclasses of the channels defined by (\ref{chandef}).
  \begin{thm}  \label{thm:comm-evec}
Let $\Phi$ be a channel of the form \eqref{chandef}  for which all of
the unitary operators $V_k$ have a common eigenvector $|\psi\ket$.  Then for
any CPT map $\Omega$
 \be  
  \hbox{a)} \qquad  & \qquad  &  \norm{ \Phi(\proj{\psi} )}_p =    \nu_p(\Phi) = \nu_p(\Gamma_a^\dep)   
     \quad \forall ~~ p  \geq 1  \label{main}  \\
   \hbox{b)} \qquad  & \qquad  &   
    \nu_p(\Phi \ot \Omega) =   \nu_p(\Phi)  \nu_p(\Omega)   \quad \forall ~~ p  \geq 1  \label{pmult}  \\
   \hbox{c)} \qquad  & \qquad  &  S[\Phi(\proj{\psi} )]=  S_{\min}(\Phi) =   S_{\min}(\Gamma_a^\dep)  
     \\
  \hbox{d)} \qquad   & \qquad  &    
     S_{\min}(\Phi \ot \Omega) =   S_{\min}(\Phi)  + S_{\min}(\Omega)  \qquad \qquad    \label{Sadd2}
\ee  
 \end{thm}
\pf First, observe that  
 \be  \label{dep.mult}
     \Phi(\rho) & = &  \sum_k \tfrac{a_k}{a} V_k  \bigg[ a \rho  + (1-a) (\tr \rho) \, \tfrac{1}{d} I \bigg]V_k^{\dg} 
         \\
           & = &  \sum_k \tfrac{a_k}{a} V_k  \Gamma_a^\dep(\rho)  V_k^{\dg} \nn
 \ee
is a convex combination of conjugation with $V_k$ composed with the depolarizing 
channel.   Therefore, for any density matrix $\rho$
\be  \label{comp_nup}
   \norm{ \Phi(\rho) }_p  & \leq  & \sum_k \tfrac{a_k}{a} \, 
         \norm{ V_k  \Gamma_a^\dep(\rho)  V_k^{\dg}}_p  \nn \\
            & \leq  &  \sum_k \tfrac{a_k}{a} \, \nu_p(\Gamma_a^\dep)  = \nu_p(\Gamma_a^\dep).
\ee   
Now consider $\rho = \proj{\psi}$ with $|\psi \ket$ the common eigenvector of $V_k$.  Then
\bee
\norm{\Phi\big(  \proj{\psi} \big) }_p= \norm{ a  \proj{\psi} + \tfrac{1-a}{d} I }_p =   
     \Gamma_a^\dep\big(  \proj{\psi} \big) = \nu_p(\Gamma_a^\dep),
     \eee
     where we used part (b) of Theorem~\ref{thm:dep}.      Therefore,
 $\nu_p(\Phi) $ is at least as big as $ \nu_p(\Gamma_a^\dep)$.   Combining this with      
 \eqref{comp_nup}, proves part (a).
 
 To prove (b), we proceed similarly, using \eqref{dep.mult}, to see that
 \be
  \norm{ (\Phi \ot \Omega)(\rho_{12}) }_p & \leq &    \sum_k \tfrac{a_k}{a} 
     \,   \norm{ (\Gamma_a^\dep \ot \Omega)(\rho_{12}) }_p \\
        & \leq &    \sum_k \tfrac{a_k}{a}  ~  \nu_p(\Gamma_a^\dep) \, \nu_p(\Omega) \\
        & = & \nu_p(\Gamma_a^\dep) \, \nu_p(\Omega) = \nu_p(\Phi) \, \nu_p(\Omega)
 \ee 
where the last step used part (a).   Since we can achieve  $ \nu_p(\Phi) \nu_p(\Omega)$
using a product state, this proves (b).   Parts (c) and (d) then follow by the
established  technique \cite{AHW} of taking the right derivative at $p = 1$.   \qed

By choosing all $V_k = W^k$ with $W$ a unitary matrix which generates a
cyclic group of order $d$, one can construct channels with precisely $d$
input states whose outputs have optimal purity.   Additional channels with
$d$ states of optimal output purity are discussed in Section~\ref{sect:diag}.
Channels for which  each $V_k$ has the form $\sum_{j = 1}^m \proj{f_j} \op W_k$
with $|f_j \ket$ a set of $m$ mutually orthonormal vectors and $W_k$ unitary
operators on $\big[\hbox{span}\{ |f_j \ket \}\big]^{\perp}$ are more interesting.
Several classes of examples are discussed in detail in Section~\ref{sect:exam}.
When the $W_k$ have no common eigenvectors,  it follows from  
 Theorem~\ref{thm:impure} below that these channels have
precisely $m$ mutually orthogonal states of optimal purity.
One can construct channels with $m =1, 2, \ldots d-2$;  however,   if
the $V_k$   have ${d \!-\!1}$ common eigenvectors,  then they have $d$
common eigenvectors, precluding the possibility that $m = d-1$.

\begin{thm} \label{thm:impure}
Let $\Phi$ be a channel of the form \eqref{chandef} and let
$\rho$ be any density matrix other than the projection onto a common
pure state eigenvector of all $V_k$.   Then 
    $\norm{\Phi(\rho)}_p < \nu_p(\Gamma_a^\dep)$ and
    $S[\Phi(\rho)] > S_{\min}(\Gamma_a^\dep)$.
    \end{thm}
 \pf  Under the hypothesis of the theorem,  
 \be
 \bgnorm{ \sum_k  \tfrac{a_k}{a} V_k \rho V_k^{\dg} }_{\infty} < 1
 \ee
 and one can write the eigenvalues of  $\sum_k  \tfrac{a_k}{a} V_k \rho V_k^{\dg}$
 as $[x_1, x_2, \ldots x_d]$  with $x_1 < 1$.   Then the eigenvectors of
 $\Phi(\rho)$ are
 \be
   [ax_1+ \tfrac{1-a}{d},  \, a x_2 + \tfrac{1-a}{d}, \ldots a x_d + \tfrac{1-a}{d}]
      \prec [a + \tfrac{1-a}{d},   \tfrac{1-a}{d}, \ldots  \tfrac{1-a}{d}] .
 \ee
Thus, the eigenvalues of $\Phi(\rho)$ are majorized by those of
$\Gamma_a^\dep(\proj{\psi}$ for any pure input $|\psi \ket$.  \qed

\begin{thm}  \label{thm:m-evec}
Let $\Phi$ be a channel of the form \eqref{chandef} for which the
unitary operators $V_k$ have precisely $m$ mutually orthogonal common
eigenvectors with $m < d$.   Then $\rho_\av \neq \tfrac{1}{d} I$ and   
at least $(d \mm m)$  states in the optimal input ensemble have
$S[\Phi(\rho_i)]  > S_{\min}(\Phi)$.
\end{thm} 
\pf When the number of common eigenvectors $m < d$ , it follows that
one can not find a set of $d$ mutually orthogonal pure inputs $\rho_i$ for which
$S[\Phi(\rho_i)] = S_{\min}(\Phi)$.   Therefore, one can not find an input ensemble
such that both 
$\sum_i \pi_i \rho_i = \frac{1}{d} I$ and $S[\Phi(\rho_i)] = S_{\min}(\Phi) ~ \forall \, i$
hold.
Therefore, we must have
\be  \label{Ctest}
  C_{\hv}(\Phi) < \log d - S_{\min}(\Phi).
\ee 
Since 
\be
   \sup_{\omega \in \cd} H[ \Phi(\omega), \, \Phi(\tfrac{1}{d} I )] = 
         \log d - \inf_{\omega \in \cd} S[\Phi(\omega)] =  \log d- S_{\min}(\Phi),
\ee
it follows from \eqref{Ctest} and  \eqref{maxmin} that  $\frac{1}{d} I$ is not the optimal average 
input.   

If we know that the optimal signal ensemble has at least $d$ inputs, then
at least $d - m$ of them must satisfy $S[\Phi(\rho_i)]  > S_{\min}(\Phi)$.
\qed

Although we are primarily interested in channels which are trace-preserving,
multiplicativity results, e.g., \eqref{pmult} can often be proved using only the CP condition.
Moreover, Audenaert and Braunstein \cite{AB}  showed that multiplicativity
of a special class of CP maps would imply  superadditivity of entanglement of
formation.   Therefore, we notice  that a weaker version of
Theorem~\ref{thm:comm-evec}  can be extended
to maps of the form \eqref{dep.mult} in which the $V_k$ are contractions rather
than unitary, i.e. $V_k V_k^{\dg} \leq I$.
 \begin{thm}   \label{thm:contract}
Let $\Phi$ be a CP map of the form 
\be
    \Phi(\rho) = \sum_k \tfrac{a_k}{a}  V_k \Big[ a \rho +  (1-a) ({\rm Tr} \rho)\tfrac{1}{d} I \Big] V_K^{\dg}
\ee  for which all of
the operators $V_k$ are contractions with a common eigenvector $| \psi \ket$
satisfying $V_k | \psi \ket = e^{i \theta_k } | \psi \ket$ .  Then for
any CP map $\Omega$, \eqref{main}--\eqref{Sadd2} hold.
\end{thm}
 \pf   The assumption that the eigenvalues of the common eigenvector have
  $| e^{i \theta_k }| =1$ implies that   $ \nu_p(\Phi) $ is at least as large as
$\nu_p(\Gamma_a^\dep) $.    For any contraction $V$, the eigenvalues 
of $VAV^{\dg}$ are submajorized by those of $A$, which we write as
$[\alpha_1, \alpha_2 \ldots \alpha_d]$.   To see this, write 
$A = U A_D U^{\dg}$
with $U$ unitary and $A_D$ the diagonal matrix with elements
$\delta_{jk} \alpha_j$.   Then $X = VU$ is also a contraction and the
diagonal elements of $VAV^{\dg}$ are $\sum_j |x_{ij}|^2 \alpha_j$
which are submajorized by $[\alpha_1, \alpha_2 \ldots \alpha_d]$.
By applying this to $A = a \rho + (1-a) \tfrac{1}{d} I$, the result follows
by the same argument as before.

 \pagebreak

\section{Diagonal $V_k$}  \label{sect:diag}

Before discussing several types of asymmetric channels, we consider 
channels for which  all $V_k$ are simultaneously  diagonal, as well as unitary.   
This includes the case $V_k = W^k$,
with $W^d = I$, mentioned earlier.    In all these situations, one has
precisely $d$ states of minimal output entropy and the capacity is
\be  \label{caplogd}
    C_{\hv}(\Phi) = \log d - S_{\min}(\Phi)  = \log d - S_{\min}(\Gamma_a^\dep).
\ee
It then follows  from the additivity of $S_{\min}(\Phi)$ in part (d) of 
Theorem~\ref{thm:comm-evec}
that $C_{\hv}(\Phi)$ is also additive in the sense 
$C_{\hv}(\Phi \ot \Phi) = 2 C_{\hv}(\Phi)$.   

The channels considered in this section are, therefore, convex combinations
\be   \label{diag+}
    \Phi(\rho) = a \Phi^{\rm diag}(\rho) + (1 \mm a) (\tr \rho) \, \tfrac{1}{d} I
\ee
of the completely noisy map and a ``diagonal channel''  
of the form $\Phi^{\rm diag}(\gamma) = \sum_k a_k V_k \gamma V_k^{\dag}$
with $a_k > 0$.   The term diagonal channel was introduced
 by King  \cite{King5} for CP maps whose Kraus operators are
simultaneously diagonal.   King also showed that
$\Phi^{\rm diag}(\gamma) = B*\gamma$ where $*$ denotes the Hadamard
 product , $B$ is  a positive semi-definite matrix, and $\gamma$ is written in
 the basis in which the $V_k$ are diagonal.    When 
 $V_k$ is unitary, its diagonal elements can be written as 
 $e^{i\phi_{km}},  ~m =1, 2 \ldots d$ and
$   b_{mn} = \sum_k a_k e^{i(\phi_{km} - \phi_{kn})}$.
If one also requires $\Phi^{\rm diag}$  to be
 trace-preserving,  then $\sum_k a_k = 1$ and $b_{mm} = 1 ~ \forall \, m$.
 This implies that the states $\proj{m}$ are fixed points of
 $\Phi^{\rm diag}$ so that it has $d$ pure state outputs. 
Hence additivity of both minimal output entropy and Holevo capacity
hold trivially for diagonal CPT maps.

In the examples \eqref{diag+} considered here, the corresponding outputs are
  $\Phi(\proj{m}) = a \proj{m} + (1 \mm a) \tfrac{1}{d} I $, $m = 1,2, \ldots d$
 which  yield $d$ states of minimal output entropy.     As noted above, this
 implies, that they satisfy  \eqref{capd} and \eqref{Cadd} when
 $\Omega = \Phi$.   Since Theorem~\ref{thm:comm-evec}
 holds, \eqref{main}--\eqref{Sadd2} are also satisfied.

 The depolarizing channel, \eqref{dep}, \ satisfies
  the general covariance condition 
  $\Phi(U \rho U^\dag)  \linebreak = U \Phi(\rho) U^\dag$ for arbitrary unitary
  matrices $U$, but this does not extend to channels of the  form
  \eqref{chandef}.   However,  when $V_k = W^k$ with $W = U X_d U^{\dag}$ and $U$ 
unitary, the channel  satisfies the weaker condition \eqref{Hvcov}
using the generalized Pauli matrices  $U X_d^m Z_d^n U^{\dag}$.

 Note that $W = U X_d U^{\dag}$ is equivalent to the assumption 
 that $W$ has eigenvalues $e^{i 2 \pi m/d}, ~ m = 0, 1 \ldots d \mm 1$. 
 However, one can have a unitary $W$ with $W^d = I, ~W^m \neq I, ~ m < d $
 but $W \neq U X_d U^{\dag}$.  For example, with $d = 5$, 
 choose $W$ to have eigenvalues $e^{i 2 \pi/5},  e^{i 2 \pi /5}, e^{i 2 \pi 3/5},  1, 1$.
 
 More generally, of course, one could choose $V_k$ with eigenvalues
 $e^{i\phi_{km}}$ without any rational relationship between eigenvalues
 for a single $V_k$ or between those for $V_j$ and $V_k$.    Then 
  \eqref{capd} still holds, despite the absence of any obvious group
  for which \eqref{Hvcov} holds.   However, we can not completely
  exclude the possibility of a hidden group.

\section{Asymmetric examples}  \label{sect:exam}
  
\subsection{Qutrit channels}  \label{sect:qutrit}

We will now study in detail the case $d = 3$, with
\be
V_k =  e^{i \theta} \proj{e_0} \op \sigma_k  = 
  \pmx e^{i \theta} & 0  \\ 0 & \sigma_k \emx, \quad k \in \{0,1,2,3\}
  \ee
with the convention that $\sigma_0 = I$.    As discussed in Appendix~\ref{app:qubit}
we can assume   that $a_0 \geq a_1$.

 It follows from Theorems~\ref{thm:comm-evec} and \ref{thm:m-evec}  that $\Phi$
 has exactly one state of minimal output entropy $\proj{e_0}$ and two orthogonal states
  $\proj{e_{\pm}} = \half[I \pm \sigma_1]$  whose outputs have eigenvalues 
$[a \frac{1+\lambda_1}{2} + \frac{1-a}{3}, a \frac{1-\lambda_1}{2} + \frac{1-a}{3},  \frac{1-a}{3}]$.
Here $ \lambda_1 $ is given by \eqref{lambdak}, with $i=1$.
If these states are the optimal inputs $\rho_j$, symmetry implies that the optimal average
input has the form
\be  \label{in3ex}
   \rho_\av = (1-2x)  \proj{e_0} + x \,  \proj{e_{+1}} + x  \, \proj{e_{-1}},
\label{qutritinput}
\ee
for which the optimal average output is
\be  \label{out3ex}
   \Phi(\rho_\av) = \big(a(1\!-\!2x)  +  \tfrac{1-a}{3}\big) \proj{e_0} +  \big(ax + 
      \tfrac{1-a}{3}\big)  \big( \proj{e_+}   +  \proj{e_-} \big).
\ee
We want to optimize the capacity
\be   \label{cap3ex}
  S[\Phi(\rho_\av)(x)] - \big[(1-2x)S[\Phi(\rho_0)] + x S[\Phi(\rho_{+1})] +  x S[\Phi(\rho_{-1})].
\ee
Since, $S[\Phi(\rho_{+1})] =   S[\Phi(\rho_{-1})]$, differentiating \eqref{cap3ex} gives the condition
\be
     2a \log \big( \tfrac{1+2a}{3} - 2ax \big)  -  2a \log \big( \tfrac{1-a}{3} +ax \big) =
     -2   S[\Phi(\rho_0)] + 2  S[\Phi(\rho_{\pm 1})]
\ee
or
\be  \label{t1}
    \log \frac{1 - a + 3ax}{1 +2a - 6ax} =   - \tfrac{1}{a} \Delta S
\ee
where $\Delta S =  S[\Phi(\rho_{+1})] -   S[\Phi(\rho_0)] > 0 $.  This has the solution
\be  \label{eq:3x}
   x = \frac{ (1+2a)2^{-\Delta S/a} - (1-a)}{3a \big(1 + 2^{-\Delta S/a}. 2\big)}.
\ee
It is easy to verify that $x < \frac{1}{3}$ confirming the intuition that the optimal
input will be shifted toward the state $|e_0\ket $.

Let $\rho_{x}$ denote the average for the ensemble corresponding
to the optimal $x$ \eqref{eq:3x} and $C_\hv^x(\Phi)$ the corresponding
capacity \eqref{cap3ex}.      To show that $\rho_{x}$ is the true optimal average which yields  
 $C_\hv(\Phi)$, we need to verify that
 $H[\Phi(\omega), \Phi(\rho_{x})]  \leq  C_\hv^x(\Phi)  $ for all choices of 
 $\omega$.  This has
 been done numerically for a large range  of $a$ and $\lambda_1$.
 
 \subsection{Doubly depolarizing channels}   \label{sect:qu4}
 
 We introduce some notation.  Let  $\{\proj{e_j} \}$ be an orthonormal basis for
 ${\bf C}^d$, $E_m$ the projection on span$\{ |e_1 \ket, |e_2 \ket \ldots |e_m \ket \}$,
 and $E_m^\perp$ is the projection on the orthogonal complement
  span$\{ |e_m \ket, |e_{m+1} \ket \ldots |e_d \ket \}$
   
 Now suppose that $\Phi$ is a channel of the form \eqref{chandef} in
which each $V_k$ has the
 form  $V_k = E_m \op W_k = \pmx  E_m & 0 \\ 0 & W_k \emx$ where the $W_k$ are chosen
 to be unitary $(d \mm m) \times (d \mm m)$ matrices such that on $E_m^{\perp}{\cal H} $
 \be
    \sum_k \tfrac{a_k}{a} W_k \rho W_k^{\dag} = b \rho + 
       (1-b) \,  (\trp_{E_m^{\perp} {\cal H}} \, \rho) \, \tfrac{1}{d-m} E_m^{\perp}.
\label{bdef} 
\ee
It suffices to choose $W_k$ to be the generalized Pauli matrices defined before
\eqref{noisI} and let  $a_k = a(1-b)/(d-m)^2$ for all $k$ except  
$a_0 = a [b (d-m)^2 + (1-b)]/(d-m)^2 $.
For the case $d=4$ and $m=2$, this reduces to $W_k = \sigma_k$ with 
$a_0 = a(3b+1)/4$ and $a_j =  a(1-b)/4$ for $j =1,2,3$.

 The action of $\Phi$ is similar to a depolarizing channel when restricted to
 $E_m {\cal H} $ or $E_m^{\perp}{\cal H} $.  More precisely,
 \begin{align}   \label{doub}
    \Phi(\proj{e}) & =    a \proj{e} + (1-a)   \tfrac{1}{d} I  & \forall ~|e\ket \in  E_m {\cal H}  \\
     \Phi(\proj{f}) & =    ab \proj{f} + a(1-b) \tfrac{1}{d-m} E_m^{\perp} +(1-a)  
       \tfrac{1}{d} I  &  \forall ~|f\ket \in  E_m^\perp {\cal H} 
 \end{align}
 The case $m = 1$, $d = 3$ is a special case of the  channels in the preceding section.

 We expect that capacity can be achieved by a (non-unique) ensemble  
 with $d$ inputs consisting of $m$ orthogonal vectors in $E_m {\cal H}$ 
 and $d-m$  orthogonal vectors in $E_m^\perp {\cal H}$.  (There is no
 loss of generality in assuming that  the optimal inputs
 can be written as  $\rho_j = \proj{e_j}$.)   By symmetry the probabilities
  for such an optimal ensemble satisfy  $ \pi_j =
     \begin{cases}   t & \text{for} ~   j \leq m \\   t^{\perp} & \text{for}  ~  j > m
     \end{cases} ~ $ 
  with $mt + (d-m) t^{\perp}  = 1$.   Thus 
\linebreak $\rho_\av = t E_m + t^{\perp}  E_m^{\perp}$  and 
 \be  \label{outDDex}
   \Phi(\rho_\av) = atE_m + a t^{\perp} E_m^{\perp} + (1-a) \tfrac{1}{d} I,
 \ee
 so that $C_\hv(\Phi)$ is the result of optimizing
 \be   \label{cap:DDex}
    S(  \Phi(\rho_\av) ) - mt S[\Phi(\proj{e_1})] - (d-m) t^{\perp} S[\Phi(\proj{e_d})] .
 \ee
One finds that the optimal $t$ satisfies
\be  \label{t.opt}
 a \log \frac{ad t^{\perp}  + 1-a}{adt + 1-a} = - \Delta S
\ee
where $\Delta S = S[\Phi(\proj{e_d})] - S[\Phi(\proj{e_1})] > 0$.  This implies that,
as expected, the solution will have $t > \tfrac{1}{d} > t^{\perp}$.  
It also agrees with \eqref{eq:3x} when $d =3, m = 1$ and $x = t^{\perp}$.
 When  $d = 2m$, \eqref{t.opt} has the solution
\be   \label{tperp_d=2m}
     t^{\perp} = \frac{1}{ad} \frac{a(1 + 2^{-\Delta S/a}) -  (1 - 2^{-\Delta S/a}) }{1 + 2^{-\Delta S/a}}.
\ee


 \subsection{Successively depolarizing channels}  \label{sect:qugen}
 
 The next example generalizes the qutrit case in a different way.
 We now choose $V_k = E_1 \op W_k   $ with $m = 1$ so that 
 \be
    \sum_k  a_k V_k \rho V_k^{\dg} = a \bigg[  E_1 \rho E_1  \op \bigg(\sum_k b_k W_k E_1^{\perp} \rho  E_1^{\perp}W_k^{\dg} \bigg)
        + (1-b)  (\tr E_1^{\perp} \rho) \, \tfrac{1}{d-1} E_1^{\perp} \bigg] 
 \ee
with $\sum_k b_k = b$.   Equivalently,
 \be
    \Phi(\rho) & = &  a  E_1 \rho E_1  +    \\
     \nn    &  & + 
          \sum_k  a b_k  W_k   E_1^{\perp} \rho  E_1^{\perp} W_k^{\dg} + 
    a(1-b) \,  (\tr E_1^{\perp} \rho) \, \tfrac{1}{d-1} E_1^{\perp} +  (1-a) (\tr \rho) \tfrac{1}{d} I
 \ee
 Proceeding in this way, we can inductively construct a channel with the property
 that the input states $\proj{e_j}$ have strictly increasing output entropies,
 with each minimal when $\Phi$ is restricted to states on $E_{j-1}^{\perp}$,
 except that the last pair have equal entropy, i.e.,
 $S[\Phi(\proj{e_{d-1}})] = S[\Phi(\proj{e_{d}})] $.
 
 We now make a change of notation so that  $x_1 = \sum_k a_k, x_2 = \sum_k b_k$, etc.
 Then
 \bee
   \Phi:  \proj{e_1} ~  & \mapsto &    x_1 \proj{e_1} + \frac{1 - x_1}{d} I \\
     \proj{e_2}  ~ & \mapsto &    x_1 x_2 \proj{e_2} +  
          x_1     \frac{1 - x_2}{d-1} E_1^{\perp} +
         \frac{1 - x_1}{d} I \\
         \vdots  \quad  & & \quad \vdots \\
     \proj{e_m}  & \mapsto &   \prod_{j=1}^m x_j   \proj{e_m} +  
         \prod_{j=1}^{m\mm 1} x_j     \frac{1 \mm  x_m}{d\mm m \pp 1} E_m^{\perp} +  
         \ldots    + \frac{1 - x_1}{d} I \\
          \vdots  \quad  & & \quad  \vdots \\
       \proj{e_{d-1}}  & \mapsto &     
\prod_{j=1}^{d-1} x_j   \proj{e_{d-1}} +
           \prod_{j=1}^{d-2} x_j  \frac{1 \mm x_{d-1}}{2}E_{d-1}^{\perp}  
           \\ & & \qquad  + 
              \prod_{j=1}^{d-3} x_j  \frac{1 \mm x_{d-2}}{3} E_{d-2}^{\perp}
 + \ldots  +
                 \frac{1 \mm  x_1}{d} I \\
        \proj{e_d}  ~ & \mapsto &     \prod_{j=1}^{d-2} x_j  (1 \mm x_{d-1}) \proj{e_{d}}
          +  \prod_{j=1}^{d-2} x_j   \frac{x_{d-1}}{2}  E_{d-1}^{\perp} 
              \\ & & \qquad  +  
            \prod_{j=1}^{d-3} x_j  \frac{1 \mm x_{d-2}}{3} E_{d-2}^{\perp} 
             + \ldots  + \frac{1 \mm  x_1}{d} I
 \eee
 
\subsection{Connection with CQ and classical channels}  \label{sect:class}

For  a channel $\Phi$  of the type considered in the preceding sections, define
 $g_{jk} =  \bra e_j | \Phi( \proj{e_k}) | e_j \ket$
 so that 
 \be  \label{CQmat}
   \Phi( \proj{e_k} ) = \sum_j g_{jk} \proj{e_j} .
\ee
Explicit expressions for the channels in Sections~\ref{sect:qu4} and
 \ref{sect:qugen} are given in Appendix~\ref{app:C}.
  The matrix $G$ is   column
stochastic, and the  ``successive'' minimal entropy outputs are
the same as for the CQ channel
 \be   \label{CQequiv2}
     \Phi_{\rm CQ}(\rho) = \sum_k  \Big( \sum_j g_{jk}   \proj{e_j}.
\Big) \tr \rho \proj{e_k}
 \ee
 
 Under the assumption that the ``successive'' minimal entropy inputs
 form a set of   optimal inputs for the Holevo capacity, the optimization
 problem for the weights   in the input ensemble $\{ \pi_m , \proj{e_m} \}$
 is the same as for the corresponding CQ channel.
 Moreover, the bistochastic matrix $G$ defines a classical channel acting
on classical probability vectors in ${\bf R}^d$.   The optimization problem
for the Shannon capacity of this channel is the same as that for the
Holevo capacity of the CQ channel \eqref{CQequiv2}.

We expect the behavior of the examples in the previous sections to be
similar to that of a qubit channel of the form
\be
   \half  [I + \bw \dtsig]  \mapsto \half \big[I + \lambda_1 w_1 \sigma_1 +   \lambda_2 w_2 \sigma_2
       + (t_3 +   \lambda_3 w_3 ) \sigma_3 \big]
\ee
with  $\lambda_3 > \lambda_2 = \lambda_1$ so that image is a football
and the only non-unital
component is a translation along the longest axis.  
For such  channels,  it is well- known \cite{FN,KR1}
that the optimal inputs for the capacity
$C_\hv$ are the orthogonal states $\half[I \pm \sigma_3]$.
and the optimal weights are determined by the corresponding classical
problem.  

If the {\em conjecture} for the examples in the preceding sections (that the
optimal inputs  are orthogonal states which correspond
to ``successive'' minimal entropy inputs) holds, then, although unital,
 they behave like the non-unital qubit channel above, i.e., they are
 closely related to  a CQ  and a classical problem with
the same probability distribution for the optimal ensemble.     This
has been verified numerically for the qutrit channels of Section~\ref{sect:qutrit}
 and the double depolarizing channels of Section~\ref{sect:qu4}.

\section{Numerical determination of capacity} \label{sect:num} 

 \subsection{Description of the algorithms}
 \label{sect:algorithm}
   Our numerical work is based on the following variant of the max--min
   principle  (\ref{maxmin}a)-(\ref{maxmin}c)   \be \label{maxmin4}
  C_\hv(\Omega)  \leq \sup_{\omega \in \cd} 
           H\big[ \Omega(\omega), \Omega(\gamma) \big]    
\ee  
with equality   if and only if $\Omega(\gamma)=\Omega(\rho_\av)$.   
The equality condition follows from the argument in \cite{SW3} 
which implies that if
$\Omega(\rho_\av) \neq \Omega(\gamma)$, then at least one of the inputs 
$\rho_j$
in an optimal signal ensemble must satisfy
\bee
H\big[ \Omega(\rho_j), \Omega(\gamma)\big]  \geq 
    C_\hv(\Omega) + H\big[ \Omega(\rho_\av), \Omega(\gamma)\big]  > 
C_\hv(\Omega) .
 \eee
Note that this also implies that the optimal average output $\Omega(\rho_\av)$ 
is unique, a fact which can be proven directly from the strict concavity
of the entropy.  This uniqueness is implicit in \cite{KNR1} and stated and 
proved 
explicitly in \cite{Shir}.   It can happen (as in the first example of 
Section~\ref{sect:othex})
that there is more than one optimal signal ensemble or optimal average
input; however, the optimal average {\em output} of a channel is always 
unique.
 
 Now suppose that we have a candidate for both the optimal average
 output $\Omega(\rho_\av^{\star})$ and an associated candidate capacity 
 $C_\hv^{\star}(\Omega)$.   
 \begin{itemize}
 
 \item[a)] If there is a state $\omega$ such that
 $ C_\hv^{\star}(\Omega) <  H\big[ \Omega(\omega), \Omega(\rho_\av^{\star}) 
\big] $
 we can conclude that the candidate  is not the true optimal average.
  \item[b)] If
 $    C_\hv^{\star}(\Omega) = \sup_{\omega \in \cd}   H\big[ \Omega(\omega), 
\Omega(\rho_\av^{\star}) \big] $
  we can conclude that we have found the true optimal average and capacity,
  at least up to the accuracy of the numerical work.    Moreover, the states 
$\omega$
  which achieve this supremum are the optimal inputs for $\Omega$.
  \end{itemize}
  
   To find the supremum in \eqref{maxmin4}, we used an algorithm
 based on an optimization
 principle of Shor \cite{shor-king} which is stated and proved as
 Theorem~\ref{thm:opt1} in Appendix~\ref{app:opt}.     
This algorithm finds relative, rather than absolute, maxima and
is applied in situations in which some relative maxima are known
(or expected) to satisfy (b) above.   Therefore, for each channel tested,
 it is necessary to use it repeatedly with multiple inputs chosen 
 to ensure that it will find a state satisfying (a) if one exists.
 
 \subsection{Numerical results}
 
 \subsubsection{Single use of channel}  \label{sect:sing}
 
 We first tested our hypothesis that the ``successive'' minimal entropy 
 states for the examples in Section~\ref{sect:exam} are optimal
 inputs for the Holevo capacity.   If this hypothesis is correct, the
 weights for the optimal ensemble are given by the optimization
 problem of Section~\ref{sect:class}.   Numerical tests were
 done only for the qutrit channels of Section~\ref{sect:qutrit}
 and the double depolarizing channels of Section~\ref{sect:qu4}
 in the case $d = 4$, $m = 2$, with parameter choices similar to
 those tested for additivity.
 
 For the qutrit case, $\Phi(\rho_\av^{\star})$ and $C_\hv^{\star}(\Phi)$
 are given by \eqref{out3ex} and \eqref{cap3ex} respectively with 
  $x$ given by \eqref{eq:3x}.    The parameters   $a_k$
were chosen so that 
$a_0 > a/2$,  and $ a_0\ge a_1 \ge a_2 \ge a_3 $
with $a = 0.5, 0.52, 0.54, \ldots 0.9 $ and for each of these
$a_0 =  a/2 + 0.05, a/2 + 0.1 \ldots$ until $a_0$ exceeds $a - 0.01$.
 For each of these pairs, we considered $a_j = (a - a_0)/3$ as well as
 a selection of parameters with  $a_1 > a_2 > a_3$.
  
  For the $d = 4$, $m = 2$ case,
$\Phi(\rho_\av^\star )$ is given by \eqref{outDDex}
and $ C_\hv^{\star}(\Phi)$ by   \eqref{cap:DDex}
with $ d = 4, m = 2$ and $t^\perp$ given by
\eqref{tperp_d=2m}.    All pairs of parameters $a$ and $b$
in the set $\{0.5,  0.55, 0.6, \ldots 0.9\}$ were tested.

The starting inputs used in Theorem~\ref{thm:opt1}
  were chosen as follows. In both cases, for each set of parameters, 
$50$ pure input states $|\psi\rangle\langle \psi|$ were  obtained
by normalizing the state
 $ |\widetilde{\psi} \rangle= \sum_{k=1}^d r_k |k\ket$ 
 where $|k\ket$  denotes the standard basis for ${\bf{C}}^d$
 and the complex coefficients $r_k$ were chosen randomly. 
  In both cases, for all choices of parameters,  
  $H[\Phi(\omega), \Phi(\rho_\av^{\star})]  \leq C_\hv^{\star}(\Phi)$
  to an accuracy of  $10$ significant figures.
  
   \subsubsection{Additivity} 
  
We tested additivity of $C_\hv(\Phi \ot \Phi)$ for 
   the channels of Section~\ref{sect:qutrit}
 and those of Section~\ref{sect:qu4}  with $d = 4$, $m = 2$.
In both cases, $\Omega(\rho_\av^{\star}) =  \Phi(\rho_\av) \ot \Phi(\rho_\av)  $
 and $C_\hv^{\star}(\Omega) = 2  C_\hv(\Phi)$ with $\rho_\av$
 and $C_\hv(\Phi)$  the expressions for a single use under the
 assumption that successively orthogonal minimal entropy inputs 
 are optimal for the capacity.     The assumption was tested 
 numerically in the  previous section.     The results of this section
 give further support for this conjecture; if it were not true, one could
 find another pair of products with capacity greater than twice
 the $  C_\hv^{\star}(\Phi)$ from the previous section.

  The algorithm in Theorem~\ref{thm:opt1}
 always yields a sequence $\omega_k$ for which \linebreak
 $H\big[ (\Phi \ot \Phi)(\omega_k), \Phi(\rho_\av) \ot   \Phi(\rho_\av) \big]  
$ in non-decreasing.
Although the limiting state  $\omega$  is stationary in the sense
of \eqref{stat}, the eigenvalue $\lambda$ need not equal the supremum in
 \eqref{maxmin4}.   Indeed, when testing additivity, 
 products of optimal inputs will always be 
 stationary  states.   Therefore,  it is important to include starting points
 which do not automatically converge to these stationary points
 if others exist.
 
 In choosing the parameters for testing additivity, it is reasonable to
exclude values for which some restriction of the channel is 
entanglement breaking (EBT).    Thus,  we focus on values
{\em well away} from the EBT regions for the corresponding depolarizing 
channel,  
i.e.,  $a \leq  0.25$ for $d = 3$ and $a \leq 0.2, b \leq \thrd$ for   $d =4$ 
in  Section~\ref{sect:qu4}. 
Similarly, for qutrits,  we choose $a_0 > \half a$.     We do 
not claim 
that channels with some EBT parameters are EBT or that 
 we can prove additivity.   However, it would be quite extraordinary
if a channel of the form \eqref{doub}  with parameters in (or near) 
the EBT regions were super-addditive
when those with larger values were not.

Because the double depolarizing examples offer possibilities for
entanglement across regions in ways not previously tested numerically,
we concentrated on this case.  For $d=4$, $m=2$, we considered 
all pairs of parameters $a,b$ in the set
$\{0.5,  0.52, 0.54, \ldots 0.98\}$.
For each pair, we used the following 
selection of input states (which are described with the convention
that $| k \ket$ denotes the standard basis in ${\bf C^4 }$):
\begin{itemize}
\item[i)] $10$ random  pure states 
$|\psi\rangle\langle \psi|$, where $|\psi\rangle$ is obtained 
by normalizing the state
$$ 
|\widetilde{\psi}\rangle 
= \sum_{j=1}^4\sum_{j=1}^4  r_{jk}  |j\rangle \otimes |j\rangle,
$$
with complex coefficients  $r_{jk}$  chosen randomly. 

\item[ii)]  $10$ maximally entangled input states $|\psi\rangle\langle \psi|$, 
where 
$$ 
|{\psi}\rangle 
=  c_1 |1\rangle \otimes |3\rangle +
  c_2 |2\rangle \otimes |4\rangle +
  c_3 |3\rangle \otimes |2\rangle +
  c_4 |4\rangle \otimes |1\rangle.
$$
with $c_k = (1/2) \exp(i \theta_k)$ and $\theta_k$
chosen randomly in $[0,2 \pi]$.

\item[iii)] 
$10$ pure input states $|\psi\rangle\langle \psi|$,
where $|\psi\rangle$ is obtained 
by normalizing the state
$$ 
|\widetilde{\psi}\rangle 
= \sum_{i=1}^4 |\phi_i\rangle \otimes |\phi_i \rangle,
$$
with each $|\phi_i \rangle $ chosen randomly as in Section~\ref{sect:sing}

\end{itemize}
For $d=3$, the same  parameter values were used as  in
Section~\ref{sect:sing} with
$30$ random  input pure states chosen  as described in (i) above.

In all the situations tested,  $C_\hv (\Phi \otimes \Phi)$ agrees with 
$2 C_\hv (\Phi)$  to $10$ significant figures.

\section{Discussion}  \label{discussion}

We have considered the effect of 
modifying a depolarizing channel by replacing $a \rho$, the 
 first term in \eqref{dep}, by different convex combinations of unitary
 conjugations.      We have shown that this leads to a rich variety
 of examples, some of which exhibit behavior previously associated
 with non-unital channels.    Nevertheless, we prove a number
 of results, including the additivity of minimal output entropy.
 
 To relate our results to other recent work, let  
 $M(\rho) = \sum_k x_k V_k \rho V_k^{\dg}$ with $x_k = \frac{a_k}{a}$
 as in (2).
 Then the channel in \eqref{chandef} can be written as  
 $\Phi = \Gamma_a^\dep \circ M$, and Fukuda's lemma \cite{F}  can be
 applied to give an alternate proof of parts (b) and
  (d) of Theorem~\ref{thm:comm-evec}.    When the $V_k$
have a common eigenvector, $M(\rho) $ has an output state of rank
one so that Fukuda's lemma   can be applied to the composition of
$M(\rho) $ with other unitarily invariant channels as discussed in \cite{F}.    In addition,
the channel $T(\rho) = \tfrac{1}{d-1} \big[ (\tr \rho) I - M(\rho) \big]$
has an output which is a multiple of a projection.  Therefore,
the results of Wolf and Eisert \cite{WE} imply that additivity 
 \eqref{Sadd} and  multiplicativity \eqref{pmult} with $1 \leq p \leq 2$ hold for
 tensor products of channels $T(\rho)$ in the ``strong'' sense defined in \cite{WE}.
 Channels $M(\rho)$ generated from diagonal $V_k$ as in Section~\ref{sect:diag} were
 considered in  \cite{WE}; however, using the $V_k$ from the asymmetric
 examples of Section~\ref{sect:exam} to generate $T(\rho)$ via $M(\rho)$
 gives new examples.

Instead of modifying the first term  in \eqref{dep}, one could change the second 
to obtain the channel
\be  \label{gen-dep}
  \Phi(\rho) = a \rho + (1-a) (\tr \rho) \gamma
\ee
with  $\gamma$ a  fixed density matrix.    The simplest such example
is the shifted  depolarizing channel 
  $\gamma = \frac{1}{d} (1-b)I + b \proj{\psi}$,   for which additivity 
 \eqref{Sadd} and  multiplicativity \eqref{pmult} for all $p \geq 1$ have
 now been proved by Fukuda \cite{F}.
However, the only results which have been proved for the general
channel  \eqref{gen-dep} are multiplicativity in the case $p = 2$  \cite{GLR}, 
and higher integers \cite{KNR2}.      Despite recent progress for special cases, 
resolving the additivity conjectures remains a challenge.

\bigskip

\noindent{\bf Acknowledgment}   This work began when MBR was a participant
in the program on Quantum Information at the Isaac  Newton Institute
at Cambridge University in 2004, and benefitted greatly from the stimulating
environment there.   ND would like to thank Daniel Oi and Alastair Kay for help with
Mathematica

\bigskip

 \pagebreak

  \appendix

  \section{Shor's optimization algorithm}  \label{app:opt}
  
  Our numerical results use the following theorem due to
   Shor \cite{shor-king}.     
   \begin{thm}  \label{thm:opt1}
  Let $\Omega$ be a CPT map and $\wh{\Omega} $ its adjoint with respect to the
  Hilbert-Schmidt inner product.   Let $\psi$ be the eigenvector corresponding
  to the largest eigenvalue of $\wh{\Omega}\big[ \log \Omega(\rho) - \log A 
\big]$.
  Then $H[\Omega(\proj{\psi}),A ] \geq  H[\Omega(\rho), A] $.
  \end{thm}
 \pf  The largest eigenvalue of $\wh{\Omega}\big[ \log \Omega(\rho) - \log A 
\big]$
 is
 \be
\lambda & =&  \sup_{\psi} \bra \psi ,\wh{\Omega}\big[ \log \Omega(\rho) - \log 
A \big] \psi \ket \\
    & =&  \sup_{\psi} \tr \proj{\psi }\wh{\Omega}\big[ \log \Omega(\rho) - 
\log A \big] 
 \ee
where the supremum is over vectors $\psi$ with $\norm{\psi} = 1$.
Let $\gamma = \proj{\psi }$ for the vector which attains this supremum.
Then
\be  \label{supineq}
  \tr \Omega( \gamma) \,\big[ \log \Omega(\rho) - \log A \big]
    & = &   \tr  \gamma \, \wh{\Omega}\big[ \log \Omega(\rho) - \log A \big]  
\nn \\
  & \geq &   \tr  \rho  \, \wh{\Omega}\big[ \log \Omega(\rho) - \log A \big]    
\\
  & = &  H[\Omega(\rho),A ] \nn
\ee
so that
\be
\lefteqn{ H[\Omega(\gamma),A ]   - H[\Omega(\rho),A ] }  \quad
   \\ 
&=  & H[\Omega(\gamma),\Omega(\rho)]  
  +  \tr \Omega( \gamma) \,\big[ \log \Omega(\rho) - \log A \big] - 
     H[\Omega(\rho),A ]  \nn \\
      & \geq & 0    \qed
 \ee

   Given a starting $\rho = \proj{\psi_0)}$, let $\gamma_1= \gamma = \proj{\psi_1}$ be
   the eigenvector before  \eqref{supineq}, and inductively define
   $\gamma_{k+1}=  \proj{\psi_{k+1}}$ using the eigenvalue equation for $\gamma_k$.
  This    gives a sequence for which  $H[\Omega(\gamma_k),\Omega(\rho)]$ 
  increases to a stationary point $\omega$ satisfying
  \be  \label{stat}
  \wh{\Omega}\big[ \log \Omega(\rho) - \log A \big] \omega = \lambda \, \omega.
  \ee

\section{Qubit channel details} \label{app:qubit}

It was shown in \cite{KR1} that any unital qubit channel can be written  as
\be  \label{qubit0}
\Phi(\rho) =    V \Big[ \ds{\sum_{k = 0}^3}
     \alpha_k  \, \sigma_k \big( U \rho U^{\dg} \big) \sigma_k \Big] V^{\dg}
 \ee
with $U, V$ unitary , the $\alpha_k >0$ with $\sum_k \alpha_k = 1$, $\sigma_0 = I$ and $\sigma_j, ~ j = 1,2,3$ the three Pauli
matrices.     There is no loss of
generality in assuming that $\alpha_0 \geq \alpha_j ~ (j=1,2,3)$; if, instead,
$\alpha_j$ is largest,  one can factor out $\sigma_j$ and rewrite $\Phi$
in the form \eqref{qubit0} with $V \raw V \sigma_j $.   Similarly, one can
choose $U,V$ to correspond to rotations in ${\bf R^3}$ so that 
$\alpha_1 \geq \alpha_j ~ (j= 2,3)$.   Finally, since the only effect of
$U, V$ is to make change of bases which have no effect on the minimal output 
entropy or the Holevo capacity,  we can assume that $U = V = I$.  Thus,
there is no loss of generality in assuming  
that $\Phi$ has the form \eqref{qubit1} with  
$   \alpha_0 \geq \alpha_1 \geq  \alpha_j  ~~ j=2,3$.   If, in
addition, $   \alpha_0 > \half$, the channel is {\em not} EBT \cite{EBT2}.
Thus, we often assume that
\be   \label{q2ord}
  \alpha_0 > \half \geq \alpha_1 \geq  \alpha_j   \quad  j=2,3.
\ee

The parameters $\alpha_k$, $k= 0,1,2,3$ and $\lambda_i$, 
$i= 1,2,3$,  in \eqref{qubit1} and 
\eqref{qubout}  are related by the conditions
\be
   1 & = &  \alpha_0 +  \alpha_1 + \alpha_2 +  \alpha_3 \\  \label{lambdak}
   \lambda_i & = & \alpha_0 +  \alpha_i - \alpha_j -  \alpha_l 
= 2(\alpha_0 +  \alpha_i) - 1
\ee
with the understanding that $i,j,l$ are distinct.    Then the input states
 $\half(I \pm \sigma_i)$ have output states $\half(I \pm \lambda_i \sigma_i)$
whose eigenvalues are
\be   \label{qeval}
  \half(1 \pm \lambda_i) = \begin{cases} ~ \alpha_0 +  \alpha_i \\ 
  ~  \alpha_j +  \alpha_l  =  1 - \alpha_0 -  \alpha_i  \end{cases}.
  \ee
  The image of the Bloch sphere is an ellipsoid whose axes
   have lengths $|\lambda_j|, ~ j=1,2,3$ with the  output states
above  at the ends of the axes.      Under the order assumption
\eqref{q2ord}, all $\lambda_j \geq 0$ and the states with optimal
output purity satisfy \eqref{qeval}
with $i = 1$.

  In the discussion of Section~\ref{sect:qutrit},
    $ \alpha_k = \frac{a_k}{a}$ and one uses  suitably modified
forms of equations \eqref{q2ord}--\eqref{qeval}.
 
\section{CQ matrices}   \label{app:C}
For  a channel $\Phi$  of the type considered in Section~\ref{sect:qu4},
the matrix defined in \eqref{CQmat} is given by
\be
    g_{jk}  & = & \begin{cases}
    a + \frac{1 - a}{d} & \quad j = k \leq m \\
    \frac{1 - a}{d} & \quad j \neq k, j \leq m ~ \text{or} ~ k \leq m \\
    ab + \frac{a(1-b)}{d-m} +  \frac{1 - a}{d} & \quad j = k > m \\
    \frac{a(1-b)}{d-m} +  \frac{1 - a}{d} & \quad j \neq k, j, k \leq m  ~~.
    \end{cases}
\ee
For a channel of the type considered in Section~\ref{sect:qugen} it is 
\be
    g_{jk}  & = & \begin{cases}
    \dfrac{1 - x_1}{d} & \quad k > 1, j = 1 \\
  g_{k,j-1} +    \ds{  \prod_{j=1}^{j\mm 1} } x_j     \frac{1 \mm  x_j}{d\mm j \pp 1} 
               & \quad k > j > 1 \\ ~ & ~ \\
            g_{j \pp 1,j} +  \ds{   \prod_{j=1}^j x_j }  
               & \quad k = j  < d     \\
               g_{jk} & \quad k < j  \\
          g_{d \mm 1, d \mm 1}     & \quad k = j = d \end{cases}
\ee


  \end{document}